\documentclass[twocolumn,aps,prb,showpacs]{revtex4}
\usepackage{graphicx}
\begin{document}
\title{The magnetized discharge with dust: negative and
positive space charge modes}

\author{N F Cramer and S V Vladimirov}

\address{School of Physics, The University of Sydney, N.S.W. 2006, Australia}

\begin{abstract}
The structure of a discharge across a magnetic field in a dusty plasma is analysed. The dust macroparticles are 
negatively charged, but are unmagnetized because of their high mass. The electrons are highly magnetized, and the ions 
have intermediate magnetization. This results in different transport rates of the different species across the 
magnetic field. Depending on the size of the magnetic field, and the relative charge on the different species, the 
dust grains can be the dominant current carrier. The space charge clouds near the electrodes will then be determined 
by the relative mobility of the different species. The discharge can operate in one of two modes, a positive space 
charge (PSC) mode, characterized by a strong cathode fall, and a negative space charge (NSC) mode, characterized by a 
broad anode fall. 
Features unique to the dust particles can also play a role in the structure of the discharge, such as the variable 
equilibrium charge on the grains, dependent on the local potential and species temperatures, the effect of gravity on 
the grain dynamics, and the rate of charging of the grains. The dust grains can also form an ordered structure, the 
dust-plasma crystal.
A fluid model of the different species is used to calculate the structure of the resulting discharge, incorporating 
the above effects. The transition from the PSC mode to the NSC mode as the magnetic field, pressure and dust 
properties are varied is demonstrated.

\end{abstract}

%
%
\pacs{52.80.Sm, 85.10.Jz}
\maketitle

\section{Introduction}

A Particle-in-Cell/Monte-Carlo (PIC/MC) simulation (van der Straaten et al.
1994, 1997) and semi-analytic treatment (Cramer, 1997) of the radial structure of a low pressure DC cylindrical 
magnetron 
discharge has revealed a potential and electric field structure highly dependent on the pressure
and magnetic field. As either the pressure was reduced or the magnetic field
was increased, the steady state discharge was found to exhibit a transition from a positive space charge mode (PSC), 
characterised by a strong cathode fall as occurs
in an unmagnetised glow discharge, to a negative space charge mode (NSC)
characterised by a broad anode fall. The reason for the transition to the
NSC mode is the strongly reduced (according to classical theory) transport of electrons across the
magnetic field in a low pressure, strongly magnetised plasma. These two modes of the magnetised discharge have
been discussed by Thornton and Penfold (1978). 

There is little conclusive experimental evidence for the NSC mode. The cathode fall is
always observed in planar magnetron experiments, even at very low pressures
(Rossnagel and Kaufmann 1986, Gu and Lieberman 1988). Experiments by Yeom et al. 
(1989) with a cylindrical magnetron with pressures and magnetic fields similar to 
those considered in this paper and in the PIC simulation study (van der Straaten et al.
1997) showed a distinct cathode fall and no anode fall over the entire range of 
discharge parameters. Langmuir probe measurements in a cylindrical magnetron
were also reported by van der Straaten et al. (1997) with the same discharge 
parameters as used in the simulation, but the results did not agree with the
simulation results in that no anode fall was observed at low pressures and high
magnetic fields. However, Hayakawa and Wasa (1965) reported the existence of
a stable discharge operating in what appeared to be the NSC mode. The discharge featured a broad anode fall for a 
magnetic field strength greater than 4kG, which is considerably higher than the field strength predicted by the fluid 
model and the simulations ($\approx 100$kG) for the onset of the NSC mode.
In order to explain the persistence of the cathode fall in the experimental results it would be necessary for the 
electron transport across the magnetic field, at low pressures and high magnetic field strengths, to be considerably 
higher than is predicted by the classical transport coefficients. It has been postulated (eg Sheridan and Goree 1989) 
that turbulence or nonlinear coherent modes induced by instabilities in the partially ionized
plasma in crossed electric and magnetic fields (Simon 1963) may increase the
diffusion and drift of electrons, thus increasing their effective transport
coefficients.

Dust macroparticles in a discharge are negatively charged, but are unmagnetized because of their high mass. The 
electrons are highly magnetized, and the ions have intermediate magnetization. This results in different transport 
rates of the different species across the magnetic field. Depending on the size of the magnetic field, and the 
relative charge on the different species, the dust grains can be the dominant current carrier. The space charge clouds 
near the electrodes will then be determined by the relative mobility of the different species. The two modes of the 
discharge will then be affected by the charge on, and the current carried by the dust grains.
Features unique to the dust particles can also play a role in the structure of the discharge, such as the variable 
equilibrium charge on the grains, dependent on the local potential and species temperatures, the effect of gravity on 
the grain dynamics, and the rate of charging of the grains. The dust grains can also form an ordered structure, the 
dust-plasma crystal.
A fluid model of the different species is used to calculate the structure of the resulting discharge, incorporating 
the above effects. The transition from the PSC mode to the NSC mode as the magnetic field, pressure and dust 
properties are varied is demonstrated.
 
\section{The basic equations} 
We consider a one-dimensional model of a magnetron discharge between 
two parallel plate electrodes, as shown in Figure 1.

\begin{figure}
\includegraphics[width=9cm]{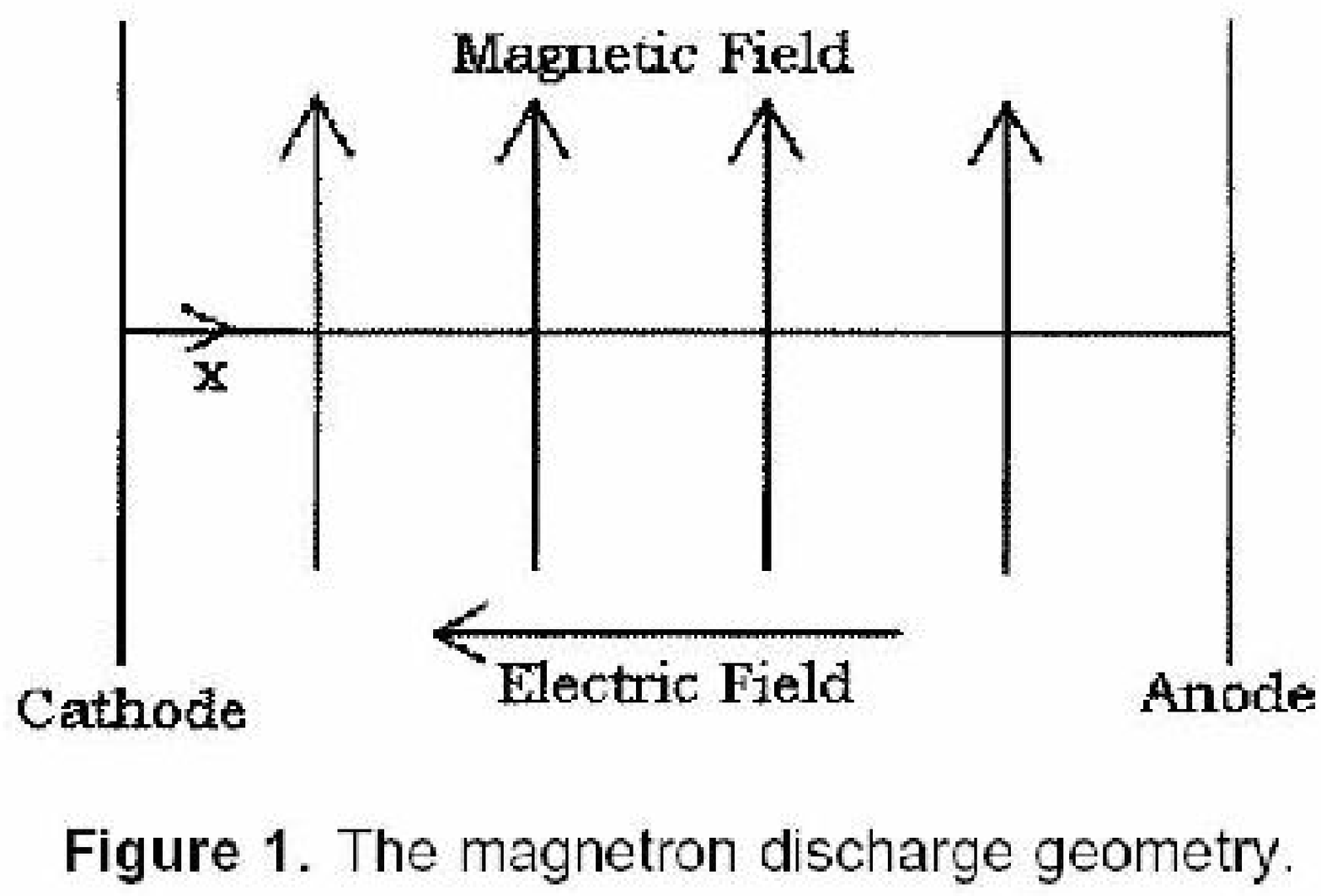}
\caption{The discharge geometry.}
\end{figure}

We assume a one-dimensional steady state distribution so that all quantities
are a function only of the distance $x$ from the cathode surface. The magnetic
field $\bf B$ is uniform and directed parallel to the cathode surface. The ions and dust grains are assumed
to be unaffected by the magnetic field, i.e. the ion and dust Larmor radii are assumed to be
much larger than the distance $d$ between the anode and the cathode.

The electrons have two components of drift velocity, $v_{Te}$ transverse to the
magnetic field in the $x$-direction, and $v_{\perp e}$ perpendicular to both the electric field $\bf E$ and to $\bf 
B$,
where

\begin{equation}
\label{vdrift}
v_{Te} =-\frac{e}{m} E \frac{\nu}{\nu^2 + \omega_c^2} ~~~~{\rm and}~~~~~
v_{\perp e} =-\frac{e}{m} E \frac{\omega_c}{\nu^2 + \omega_c^2}
\end{equation}
where $E$ is the $x$-component of $\bf E$ (negative in this case), $\omega_c$ is the electron-cyclotron frequency and 
$\nu$ is the collision
frequency of electrons with background gas atoms.
The resultant drift of the electrons is at an angle $\theta$ to the $x$-axis given by
\begin{equation}
\tan\theta = \omega_c /\nu .
\end{equation}
The basic equations used are those of Davies and Evans (1980), modified to
include the magnetic field and the dust. Thus we use Poisson's equation in one dimension, i.e.

\begin{equation}
\frac{d E}{d x} = \rho /\epsilon _0
\end{equation}
where $\rho$ is the net charge density and $\epsilon _0$ is the permittivity of
free space. This equation may be rewritten as
\begin{equation}
\frac{d E}{d x} =\frac{1}{\epsilon_0}(\frac{J_i}{v_i} +\frac{J_e}{v_{Te}}+\frac{J_d}{v_d}),
\end{equation}
where $v_i$ is the ion drift velocity in the $x$-direction (negative in
this case), $v_d$ is the dust grain drift velocity, and $J_i$, $J_e$ and $J_d$
are the ion, electron and dust current densities in the $x$-direction.

The dust current density is
\begin{equation}
J_d=n_d v_d Q_d=n_{d0}v_{d0}Q_d
\end{equation}
where we assume a constant flux of dust particles, with initial density and velocity $n_{d0}$ and $v_{d0}$. The dust 
charge varies in the discharge due to the varying local potential.

Writing the total current density as $J=J_i +J_e+J_d$, and assuming $|v_i|\ll |v_d|$, we have

\begin{equation}
\label{poisson}
\frac{d E}{d x} =\frac{J}{\epsilon_0 v_i}\left[ 1-\left( 1+ 
|\frac{v_i}{v_{Te}}|\right)j_e\right]+\frac{n_dQ_d}{\epsilon_0}                      
\end{equation}
where $j_e$ is the fraction of the total current density due to electrons. A boundary
condition that can be applied is that the electron current at the cathode is due solely
to secondary emission of electrons caused by ion impact on the cathode. The
secondary emission coefficient $\gamma =j_e/j_i$ at the cathode is assumed known.

The second basic equation we use is the electron charge conservation equation,
 or ionization avalanche equation. The electrons drift through the background
neutral gas at the angle $\theta$ to the $x$-axis and ionize the neutral gas molecules, and electron avalanches are 
formed.
These avalanches are therefore also inclined at the angle $\theta$ to the $x$-axis.
If the coordinate along this direction is $\zeta$, the normalized electron
current density in this direction is $j_{\zeta e}$ and the electric field in this
direction is $E_\zeta$, and the ionization equation may
be written

\begin{equation}
\label{jeq}
\frac{d j_{\zeta e}}{d \zeta} =\alpha j_{\zeta e},
\end{equation}
where $\alpha$ is Townsend's 1st ionization coefficient (Llewellyn-Jones 1966),

\begin{equation}
\label{alpha}
\alpha =AP \exp (-C(P/|E_\zeta | )^s)
\end{equation}
where $A$ and $C$ are constants depending on the gas, $P$ is the gas pressure and $s=1/2$ for a
monatomic gas. Since 
$\zeta =x/\cos\theta$, $j_{\zeta e}=j_e /\cos\theta$ and $E_\zeta =E\cos\theta$,
\ref{jeq} becomes

\begin{equation}
\label{current}
\frac{d j_e}{d x} = \alpha ' j_e =\frac{AP}{\cos\theta} \exp \left( -C(P/|E|\cos\theta )
                       ^{1/2}\right) j_e.
\end{equation}
The only difference in equation (\ref{current}) to the unmagnetized case is therefore the replacement of
the pressure $P$  by the ``effective pressure" $P/\cos\theta$.

The ion mobility is assumed unaffected by the magnetic field, so we assume,
as do Davies and Evans (1980), that
\begin{equation}
|v_i| = k (|E|/P)^{1/2}
\end{equation}
where $k$ is a constant. This gives a good representation of the experimental
ion drift (Ward 1962). However we note that this means that the ion
and electron drift velocities have different $E$ dependences, so $r$ is not
strictly independent of $E$ as we have assumed so far. A dependence
of $r$ on $E$ would prevent the application of the analysis used here, so
we neglect it, noting that it could cause an error in our results at high
magnetic fields.  

The charge of a (negatively charged) dust particle
is determined by the current balance equation
\begin{equation}
\label{char}
\sqrt{\pi/8}n_i(z)\bar{v}_i(z)\left[1-
\frac{2eQ_d(z)}{am_i\bar{v}_i^2(z)}\right]
=n_0v_{e}\exp\left[
\frac{eQ_d(z)}{aT_e}+\frac{e\varphi(z)}{T_e}\right].
\end{equation}

\section{Results}

The equations have been solved for a number of cases, using the above
prescription, to illustrate the
effect on the discharge of increasing the magnetic field and the density of dust particles.
The parameters  used in the numerical examples are those
of van der Straaten et al (1994), viz. $d=2.2$cm, $P=5$mTorr  and 50mTorr, and Argon
gas, for which $A=29.22{\rm cm}^{-1}{\rm Torr}^{-1}$, $C=26.6{\rm V}
^{1/2}{\rm cm}^{-1/2}{\rm Torr}^{-1/2}$, electron mobility for zero magnetic field $\mu_e=3\times 10^5{\rm cm}^2 {\rm 
Torr V s}^{-1}$ and $k=8.25\times 10^3
{\rm cm}^{3/2}{\rm Torr}^{1/2}{\rm V}^{-1/2}{\rm s}^{-1}$. 
The corresponding electron collision frequency is
$\nu = 6\times 10^9 P{\rm s}^{-1}$ where $P$ is in Torr. The ratio
of ion and electron mobilities in the unmagnetized gas is $r_0 =3.3\times
10^{-3}$ (Ward 1962).

Figure 2 shows the electric field, ion velocity and space charge
profiles for a pressure of 5mTorr. The cathode is at $x=0$ and the anode is at $x=2.2$cm. No magnetic field and no 
dust is present, and the potential drop = 209V. There is the usual positive space charge region near the cathode.

\begin{figure}
\includegraphics[width=7cm]{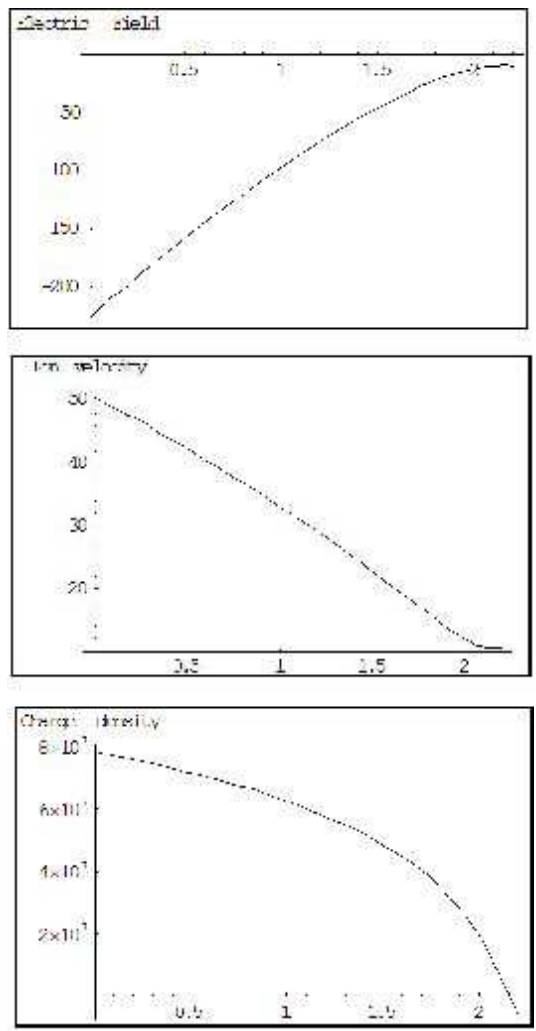}
\caption{The electric field, ion velocity and space charge
profiles for a pressure of 5mTorr. There is no magnetic field or dust.}
\end{figure}

Figure 3 shows the electric field, ion velocity and space charge
profiles. No magnetic field is present, but dust is present, with $n_{d0}=10^3$, and the potential drop = 209V. A 
negative space charge region forms near the anode, due to the dust.

\begin{figure}
\includegraphics[width=7cm]{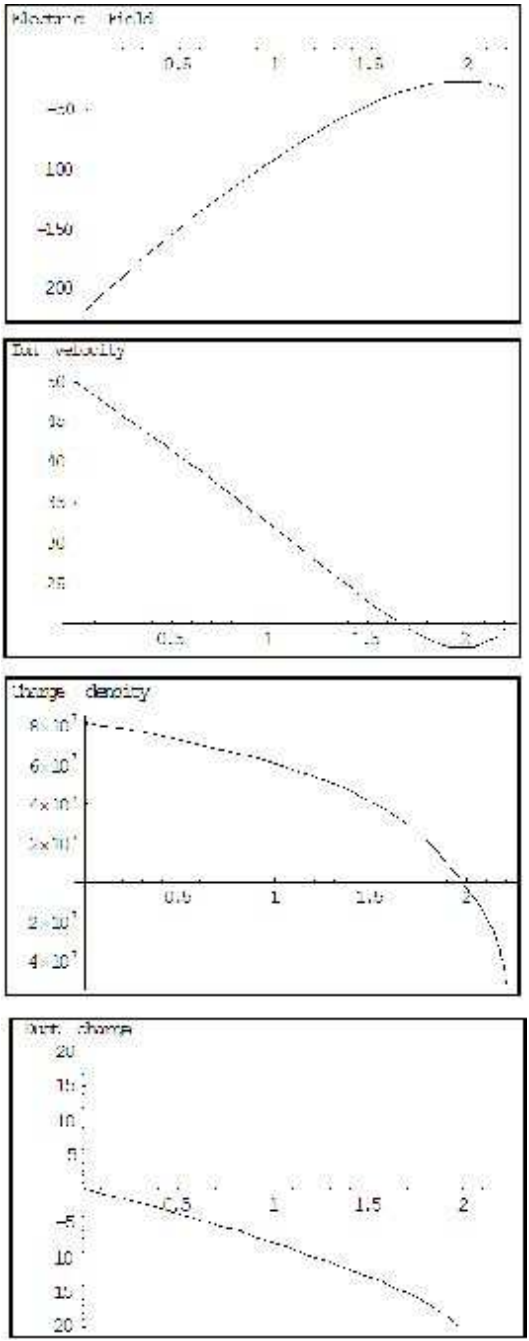}
\caption{The electric field, ion velocity, space charge and dust particle charge
profiles. There is no magnetic field, but $n_{d0}=10^3$.}
\end{figure}

Figure 4 shows the electric field, ion velocity and space charge
profiles. No magnetic field is present, but dust is present with a higher density, with $n_{d0}=3\times 10^3$, and the 
potential drop = 209V. The negative space charge region is more pronounced.

\begin{figure}
\includegraphics[width=7cm]{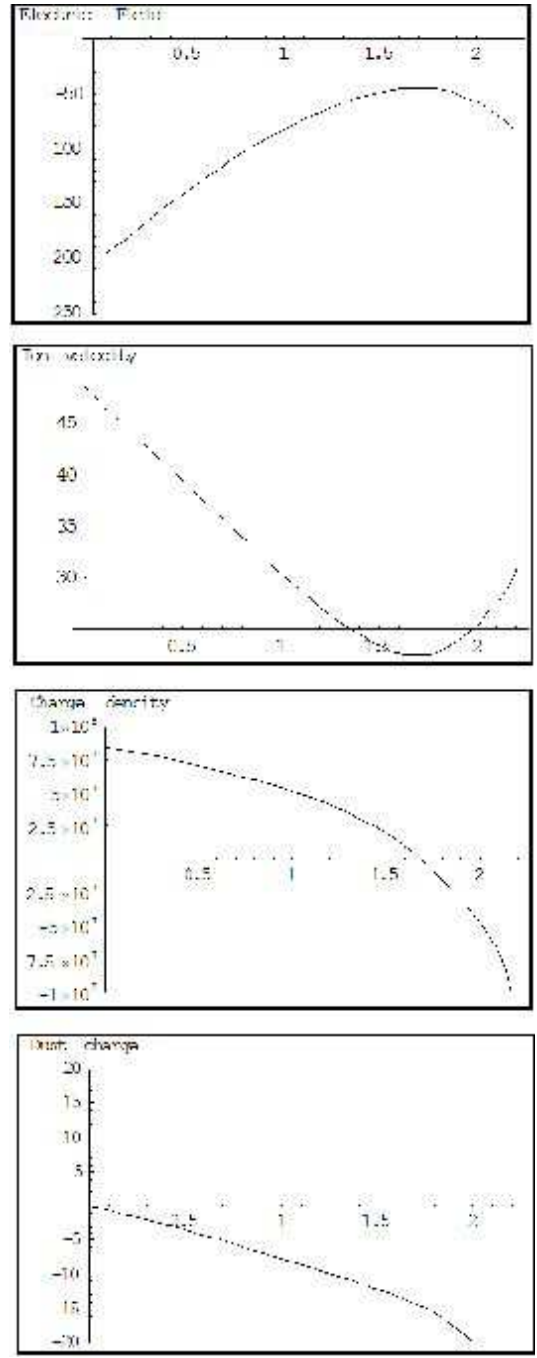}
\caption{There is no magnetic field, but $n_{d0}=3\times 10^3$.}
\end{figure}

In figure 5 a magnetic field is present, with $\cos\theta =0.2$, but no dust is present, and the potential drop = 
392V. The negative space charge region is now due to the magnetic field.

\begin{figure}
\includegraphics[width=7cm]{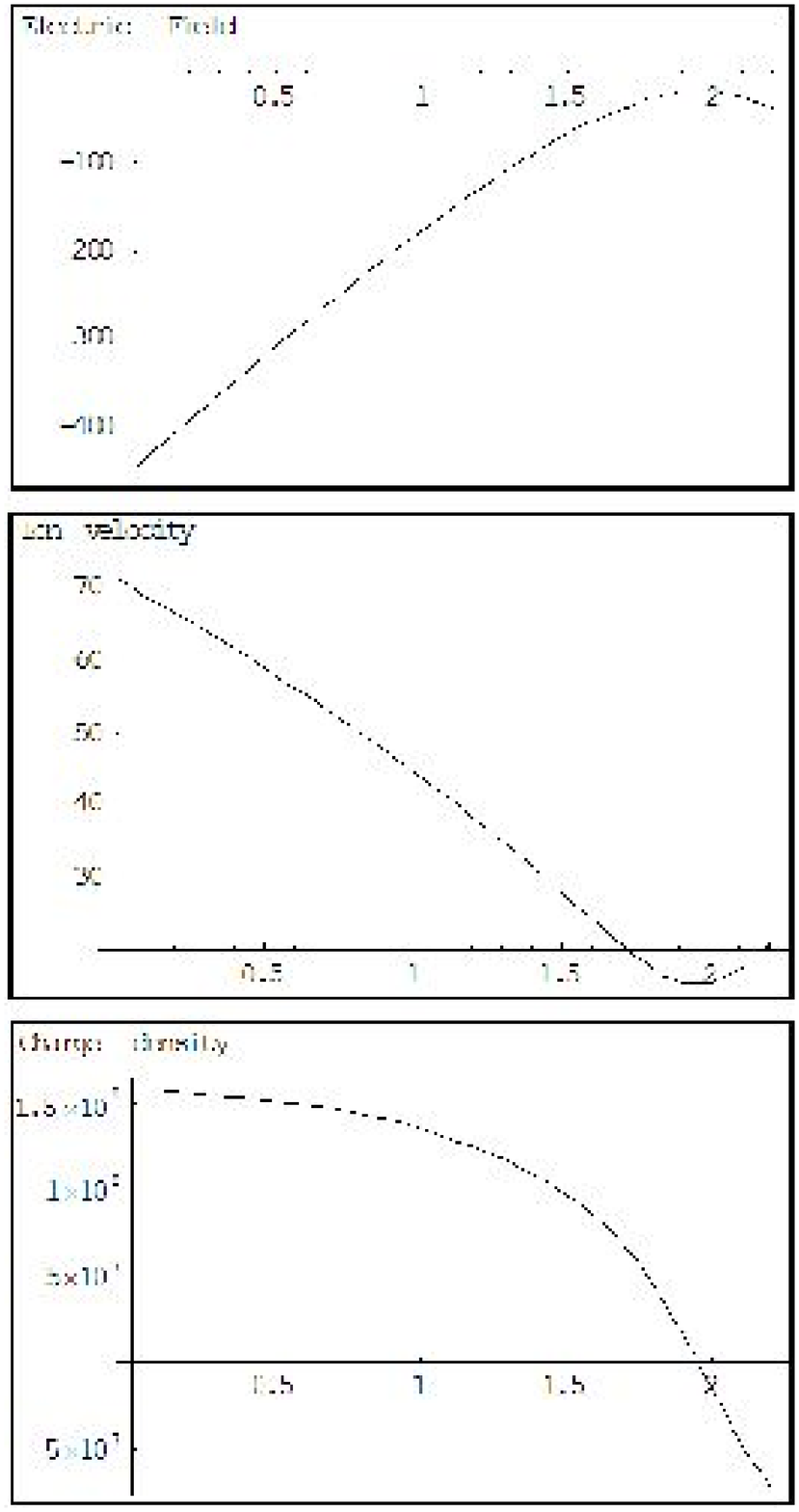}
\caption{A magnetic field is present, with $\cos\theta =0.2$, but no dust is present.}
\end{figure}

In figure 6 a magnetic field is present, with $\cos\theta =0.2$, and dust is present with $n_{d0}=10^3$, and the 
potential drop = 392V. The negative space charge region is more prominent, due to the dust.
Dust charges negative near the NSC.

\begin{figure}
\includegraphics[width=7cm]{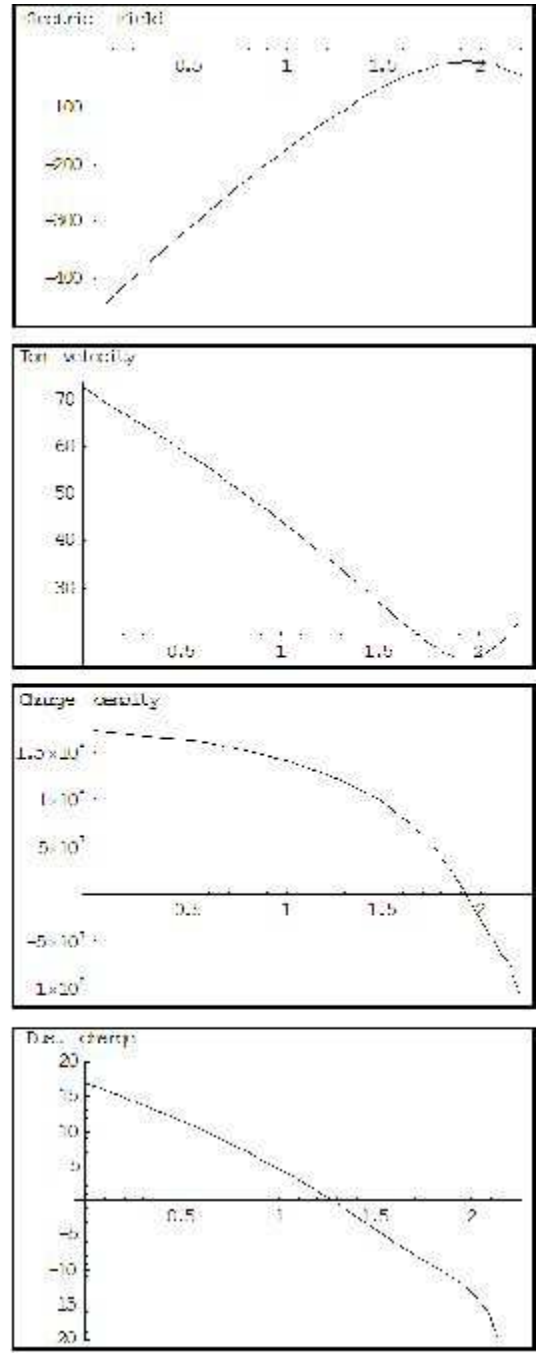}
\caption{A magnetic field is present, with $\cos\theta =0.2$, and $n_{d0}=10^3$.}
\end{figure}

\newpage

In figure 7 a stronger magnetic field is present, with $\cos\theta =0.5$, and dust is present with $n_{d0}=10^3$, and 
the potential drop = 524V. The negative space charge region is wide.
Dust charges positive over most of the discharge.

\section{Discussion and Conclusions}

A numerical solution of the electron, ion and dust fluid transport
equations for a magnetized discharge has been developed, building on
previous work for an unmagnetized steady-state glow discharge. Understanding
the transport of charged particles across the magnetic field is important
for modelling the operation of magnetron devices used in plasma processing
for industry. The effects of charge on the dust particles on the transition
from a positive space charge mode to a negative space charge mode as the
magnetic field is increased or the pressure is reduced has
been demonstrated. 

The presence of dust can create a negative space charge region near the anode, which enhances or mimics the effect of 
a magnetic field. If however the field is so strong that the ions are magnetized (future work), the dust grains may 
carry most of the current, which will enhance the positive space charge region.

\section{Acknowledgements}
The work was supported by the Australian Research Council.

\newpage

\section*{References}

\noindent
Cramer N F 1997 {\it J. Phys. D: Applied Physics}, {\bf 30}, 2573-2584

\noindent
Davies A J and  Evans J G  1980 {\it J. Phys. D: Applied Physics} {\bf 13} L161

\noindent
Gu L and Lieberman M A 1988 {\it J. Vac Sci. Technol.} A{\bf 6} 2960

\noindent
Hayakawa S and Wasa K 1965  {\it J. Phys. Soc. Japan} {\bf 20} 1692

\noindent
Llewellyn-Jones F 1966 {\it The Glow Discharge} (London: Methuen)

\noindent
Neuringer J L 1978 {J. Applied Phys.} {\bf 49} 590

\noindent
Rossnagel S M and Kaufman H R 1986 {\it J. Vac Sci. Technol.} A{\bf 4} 1822

\noindent
Sheridan T E and Goree J 1989 {\it J. Vac Sci. Technol.} A{\bf 7} 1014

\noindent
Simon A 1963 {\it Phys. Fluids} {\bf 6} 382

\noindent
Thornton J A and Penfold A S 1978 {\it Thin Film Processes}
(New York: Academic Press), eds. J L Vossen and W Kern.

\noindent
van der Straaten T A and Cramer N F 1997 {\it Phys. Plasmas}, {\bf 7}, 391--402 (2000).

\noindent
van der Straaten T A, Cramer N F, Falconer I S and James B W 1994 
47th Gaseous Electronics Conference, Gaithersburg, Maryland, USA, Abstract
published 1994 {\it Bulletin of the American Physical Society} {\bf 39} 1467

\noindent
van der Straaten T A, Cramer N F, Falconer I S and James B W 1997,
{\it J. Phys. D: Applied Physics}, {\bf 31}, 177-190 (1998). 

\noindent
Ward A 1962 {\it J. Applied Phys.} {\bf 33} 2789

\noindent
Yeom G Y, Thornton J A and Kushner M J 1989 {\it J. Applied Phys.} {\bf 65} 3816

\begin{figure}
\includegraphics[width=7cm]{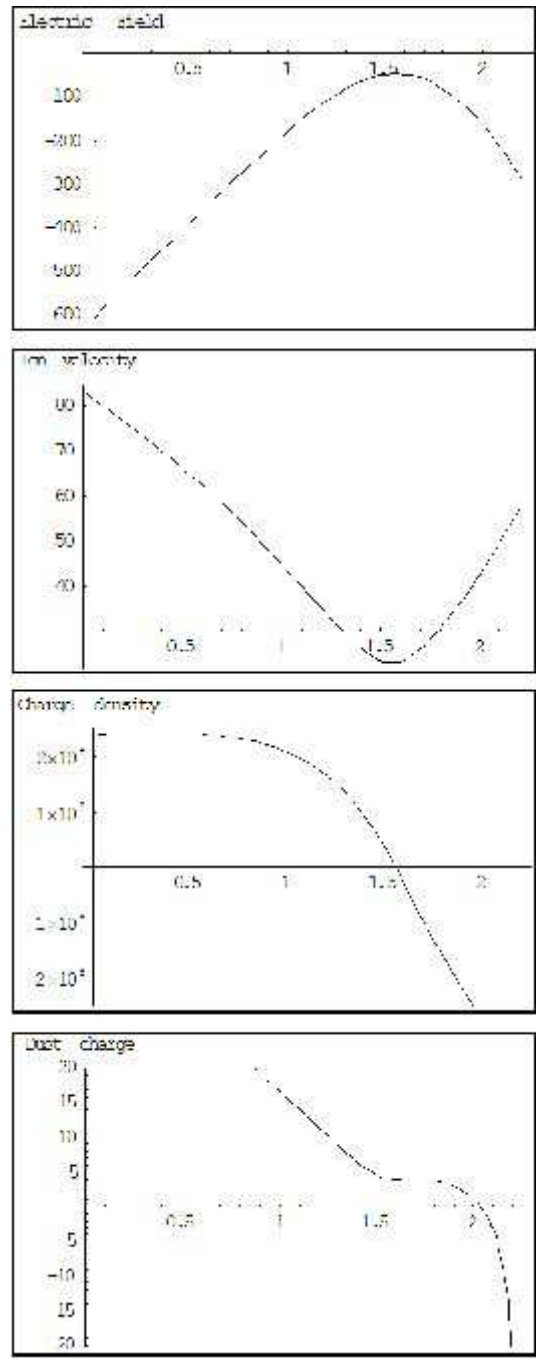}
\caption{A magnetic field is present, with $\cos\theta =0.5$, and dust is present with $n_{d0}=10^3$.}
\end{figure}

\newpage 
\end{document}